\documentclass[conference]{IEEEtran}
\usepackage{braket}

\usepackage{cite}

\usepackage{amsthm}
\usepackage{amssymb}

\usepackage[pdftex]{graphicx}

\usepackage{amsmath}
\usepackage{array}

\usepackage{xcolor}
\def\NPC{NP-Complete}
\def\NPH{NP-Hard}
\def\Xstar{\ensuremath{x^{\star}}}
\def\CCNOT{\mbox{\textbf{CCX}}}
\def\CNOT{\mbox{\textbf{CX}}}
\def\X{\mbox{\textbf{X}}}
\def\NumBits#1{%
\ensuremath{\left\lfloor\log_{2} #1\right\rfloor + 1}}
\begin{document}
%
\title{Optimization of a Quantum Subset Sum Oracle}

\author{\IEEEauthorblockN{Angelo Benoit, Sam Schwartz, and Ron K. Cytron}
\IEEEauthorblockA{Washington University\\
Saint Louis, Missouri 63130\\
\texttt{cytron@wustl.edu}}}


\maketitle

\begin{abstract}
We investigate the implementation of an oracle for the Subset Sum problem for quantum search using Grover’s algorithm. Our work concerns reducing the number of qubits, gates, and multi-controlled gates required by the oracle.

We describe the compilation of a Subset Sum instance into a quantum oracle, using a Python library we developed for Qiskit and have published in GitHub. We then present techniques to conserve qubits and gates along with experiments showing their effectiveness on random instances of Subset Sum. These techniques include moving from fixed to varying-width arithmetic, using partial sums of a set’s integers to determine specific integer widths, and sorting the set to obtain provably the most efficient partial sums.  

We present a new method for computing bit-string comparisons that avoids arbitrarily large multiple-control gates, and we introduce a simple modification to the oracle that allows for approximate solutions to the Subset Sum problem via Grover search.
\end{abstract}


%
\IEEEpeerreviewmaketitle

\section{Introduction}\label{sec:intro}

Quantum computers are attractive for solving optimization problems due to the theoretical advantage of unstructured search using Grover's algorithm~\cite{grover}.  In this paper we examine qubit-conserving implementations of an \emph{oracle}  for deployment for Grover's algorithm.  Here, the oracle of interest realizes an instance of the \emph{Subset Sum} problem (SSP)~\cite{GaJo79}. That problem is known to be \NPC{}.  While Grover cannot overcome the exponential cost currently believed to be necessary for such problems, it does asymptotically obtain quadratic speedup as compared with exhaustive search.

In particular, we are interested in the optimization problem related to SSP, which we describe in Section~\ref{sec:ssp}.  Instances of \NPC{} optimization problems, such as memory-map inference~\cite{memmap} and similar hard problems can be reduced from and to SSP.  The techniques we describe in this paper allow such hard problems to be formulated as an instance of SSP whose solution may provide good, if not optimal, solutions to the hard problems of interest.

\smallskip

The contributions of our paper are as follows.  We have developed a library to support arithmetic in an oracle (Section~\ref{sec:library}).  That library is freely available in \texttt{GitHub} and it embodies the techniques we have describe in Section~\ref{sec:qubits} that conserve quantum bits (qubits) in such oracles.  We present results of experiments conducted in \texttt{Qiskit}~\cite{qiskit} to quantify the savings of qubits for random SSP instances.  While the results we can obtain are precise under emulation, deployment on a real quantum device (such as IBM's  127~qubit quantum computers available openly for public use~\cite{ibmq}) encounters unacceptable noise.  One likely source of that noise~\cite{Wilson21} is the use of multi-controlled gates whose controlling qubits are (for example) all the bits of integers compared for equality.  In Section~\ref{sec:equality} we propose a novel method for determining the equality of two bit strings. If the strings under comparison are $m$~qubits wide, current approaches use a multi-controlled gate with $m$~control bits.  We use at most two controlling inputs for any control gate and take modest additional steps ($\Theta(\log\star (m))$) to determine equality.
Finally, we describe in Section~\ref{sec:opt} an adaptation of the oracle to find approximate solutions to Subset Sum using Grover search.

In our notation, we consistently use subscripts to denote an element taken from a collection, such as a set, and superscripts to denote a particular bit of a bit string or number.

\subsection{Subset Sum}\label{sec:ssp}

The Subset Sum problem (SSP) is a decision problem that can be described as follows.  Given a (multi)set \[S = \{a_{1}, a_{2}, \ldots, a_{n}\}\] of $n$ positive integers, is there a subset of those integers that sums to another integer, $T$? This problem is known to be \NPC{}, which, at the time of this writing, means there is no algorithm yet that can answer the above question any faster than trying all possible subsets of~$S$. While SSP may not have inherent interest, it is a common problem from which interesting, hard problems can be reduced.  One such problem we have previously considered is the \emph{memory map inference} problem. Given a sequence of storage references (loads and stores), infer which memory allocation regions waste the least amount of space~\cite{memmap}.  These would be the memory-map requests made from such a program to an operating system.

\subsection{Corresponding optimization problem}
Asking whether a memory map exists with a given efficiency can be reduced from SSP, thus showing that the memory map inference problem is \NPC{}.  More interesting and relevant to this paper, the \emph{optimization} form of SSP allows us to ask what memory map is closest to a given, specified target.  This, in turn, asks whether SSP can find a subset close to a given target~$T$.

It is important to remember that the size of an SSP instance is usually characterized by~$n$, the number of integers in set~$S$.  While that plays a role in the quantum deployment of an SSP instance, we must concern ourselves also with the integers themselves and the partial sums computed in service of finding whether a given subset sums (close) to~$T$.  Those aspects of SSP consume significant resources on a quantum computer.  This becomes apparent when we describe how to compile an SSP instance into a quantum circuit in Section~\ref{sec:grover}.
\subsection{Grover's algorithm and the oracle}\label{sec:grover}

\begin{figure}[htb]
\begin{center}
\includegraphics[width=.4\textwidth]{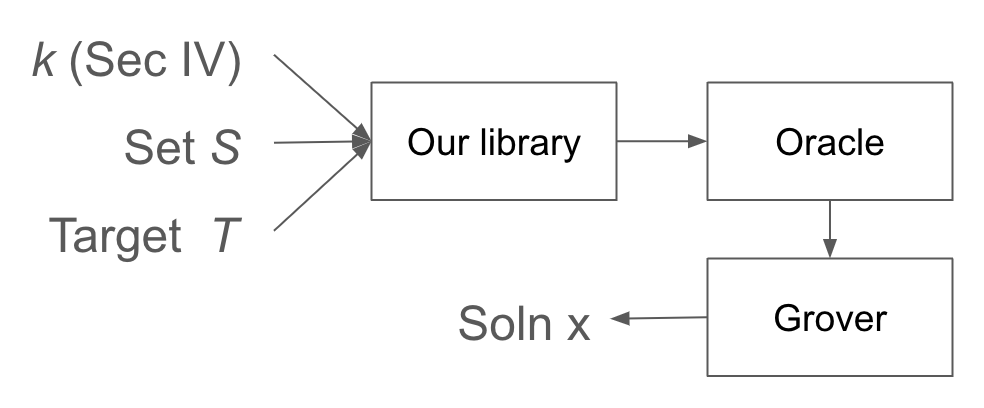}
\end{center}
\caption{Overview of the steps for generating an oracle.  $S$ is the set of integers and $T$ is the target for the sum.  The $k$ value is described in Section~\ref{sec:opt}.}\label{fig:sketch}
\end{figure}

A complete description of Grover's algorithm is beyond the scope of this paper and has been covered well elsewhere~\cite{grover,nielsen00}. 
The important aspects of that algorithm for the purposes of this paper are as follows:
\begin{itemize}
    \item The algorithm incorporates instances of an oracle that models the behavior of some function $f(x)$ where $x$ is an $n$-qubit value.  There are thus $N=2^{n}$ possible basis values for~$x$ in the computational basis. 
    \item We assume  that for exactly one such input value \Xstar{}, $f(\Xstar)=1$ and $\forall\ x \neq \Xstar, f(x)=0$. Extensions where there are more input values for which our function returns~$1$ are straightforward~\cite{grover}. 
    \item We denote the bits of $x$ as $x_{1}, x_{2}, \ldots, x_{n}$.
    \item For the purposes of SSP, $x$ encodes a particular subset of $S$ whose values might sum to $T$.  A particular bit $x_{i}$ is true if and only if the associated integer from SSP $a_{i}$ should be included in the sum.
    \item In the scenario described thus far, Grover's algorithm can find \Xstar{} in time $\Theta(\sqrt{N})$, which is quadratically faster than the exhaustive search taking $\Theta(N)$ time.
    \item Grover's algorithm performs a certain number of iterations, related to~$n$, each consisting of the oracle subjected to \emph{phase kickback}~\cite{nielsen00} followed by a \emph{diffusion} step.
    \item If Grover's algorithm requires $g$ iterations, then the quantum circuit will have $g$ instantiations of the oracle, each followed by an instantiation of the diffusion step.
\end{itemize}
We next describe how we compile an instance of SSP into an oracle for Grover's algorithm.   An outline of the steps is as follows, and a sketch of the process is depicted in Figure~\ref{fig:sketch}.
\begin{itemize}
    \item The circuit accepts $x$ ($n$~qubits) and an ancilla~$y$.  The value of $x$ selects the subset of integers from~$S$, as described above. The ancilla $y$ is initialized to the $\ket{-}$ state, as required by Grover's algorithm to induce phase kickback.
    \item For each integer $a_{i}$ in $S$, a quantum register is created and initialized to the bits of $a_{i}$.  We interchangeably call the integer and its quantum register $a_{i}$. 
    For now, each integer value in the circuit occupies 32 bits, but the improvements we suggest in Section~\ref{sec:qubits} will allocate qubits more judiciously.
    \item For each $a_{i}$, a \emph{shadow} register $b_{i}$ is created that will hold either $a_{i}$ or $0$, depending on bit $x_{i}$.  We denote bit~$j$ of a register using superscripts.  Thus we have
    \[ b_{i}^{j} = a_{i}^{j}\ \wedge\ x_{i}\]
    Bit~$j$ of the shadow register is true if and only if the corresponding bit of $a_{i}$ is true \emph{and} $x_{i}$ specifies the inclusion of integer $a_{i}$ for the sum.  The above logic is easily realized in the quantum circuit by making $b_{i}^{j}$ (initialized to $\ket{0}$) the target of a \CCNOT{} gate controlled by $a_{i}^{j}$ and~$x_{i}$.
    \item Next we must sum the shadow registers.  A running sum of the $n$ integers takes $n-1$ registers.  We do not as yet assume any order among the $n$~integers, but in Section~\ref{sec:qubits} we describe an optimal ordering for them in terms of saving qubits.
    \item The final sum~$Z$ of the chosen subset must then be compared with the target $T$.  We compare each bit $Z^{j}$ with $T^{j}$ by performing their exclusive-or using two \CNOT{} gates, as shown in Figure~\ref{fig:stdeql}(a). If each resulting bit is called $C^{j}$, then we finally determine whether all of those bits are~$0$.  In literature, this is accomplished by a multi-controlled \CNOT{} gate, whose controls are each of the bits $C^{j}$ followed by an \X{} gate and whose target then indicates whether all of the control bits are~$\ket{0}$. A diagram of this circuit is shown in Figure~\ref{fig:stdeql}(b), which includes the uncomputation after the multi-controlled gate using an~\X{} gate.  We propose an alternative computation that uses only \CCNOT{} gates in Section~\ref{sec:equality}.
\end{itemize}
The final result computed by the oracle is $\ket{1}$ if and only if the input~$x$ chooses a subset of~$S$ that sums to~$T$. Such an oracle is suitable for Grover's algorithm to find the solution~$x$.

\section{Conserving quantum bits and gates}\label{sec:qubits}

As described in Section~\ref{sec:intro}, we initially generated our oracle using 32~qubits for the integers and summation operations.  In this section we describe optimizations that allowed us to save qubits.  Experimental results from most of these ideas are reported in Section~\ref{sec:exper}, but some remain future work for us.

\subsection{Addition in a quantum circuit}

\begin{figure}[htb]
    \includegraphics[scale=0.2]{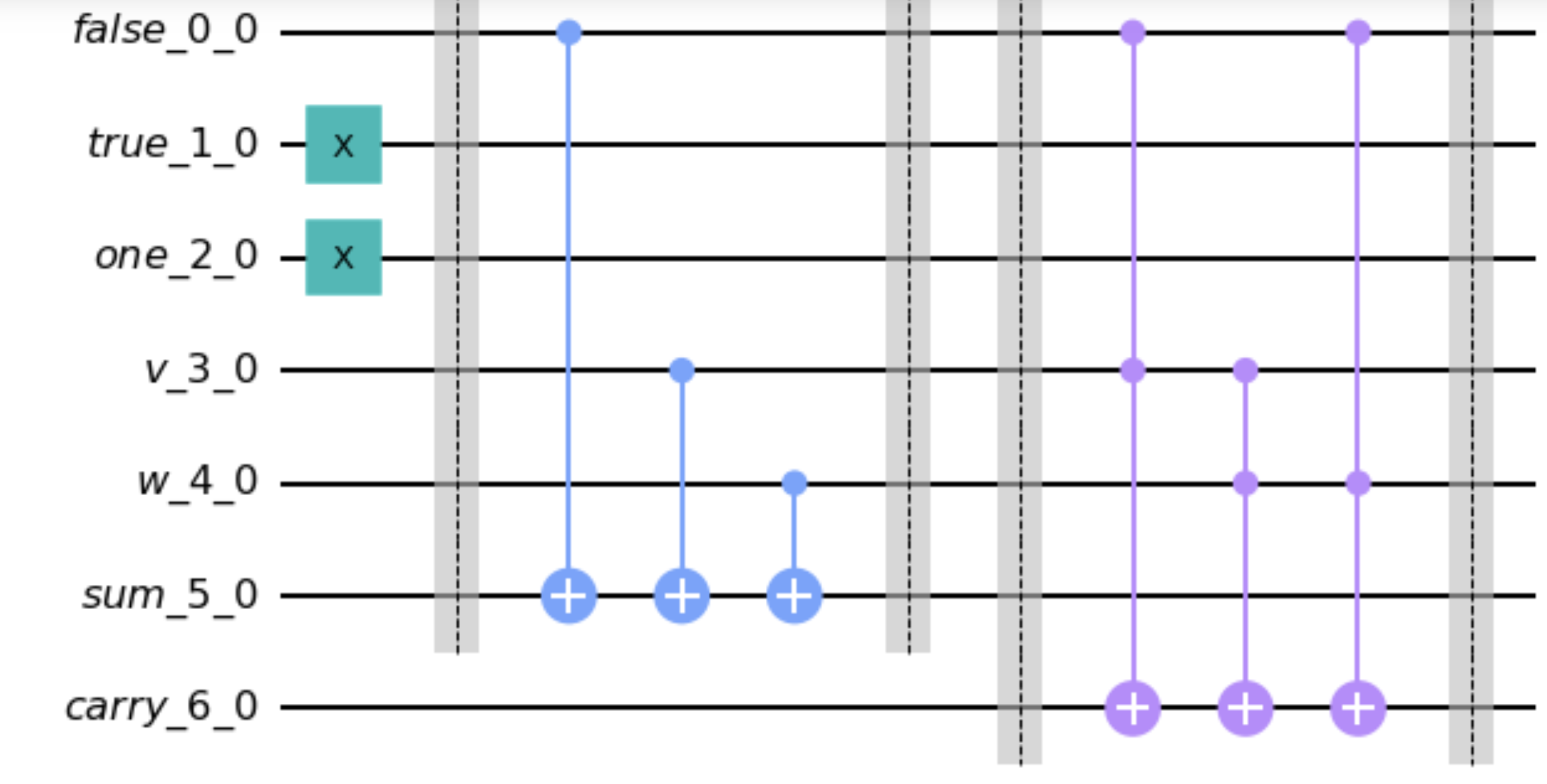}
\caption{Addition of qubits $v$ and $w$.  As this is the first addition, the carry-in is the $\ket{0}$ qubit at the top of the circuit. The circuit was generated by \texttt{Qiskit} using our library.}\label{fig:onebitadd}
\end{figure}

Our library supports the addition of integers held in quantum registers.  We mention here that we formulated the carry operation to use operations such as \CNOT{} that are readily available for quantum circuits. 
As we generate the logic to add two registers, bit-by-bit we need the sum of three qubits.  Two of the qubits, $v$ and $w$, are fresh inputs;  the third qubit is the carry $c$ from the previous bits' addition.  We compute the sum and carry(-out) out as follows:
\begin{align*}
sum &= v \oplus w \oplus c \\
carry &= vw \oplus vc \oplus wc
\end{align*}
The $carry$ expression differs from what is normally deployed for classical (non-quantum) computing, but is logically equivalent and avoids unnecessary boolean operations in a quantum circuit. 
The corresponding circuit is shown in Figure~\ref{fig:onebitadd}.

\subsection{Reducing the width}\label{sec:reducewidth}

Our results in Section~\ref{sec:exper} show that implementing all values and operations on 32~qubits integers rapidly exceeds the qubits available on real devices or even on available simulators.   The set $S$ is known \textit{a prior}, so we can compute \[ M = \sum_{i=1}^{n}\ a_{i} \]
    as the \emph{maximum} value achievable over all possible subsets for this SSP instance.
    All integers and sums could then be accommodated using
    \NumBits{M} bits.  Suppose WLOG that $a_{n}$ is the largest integer in $S$ and $S$ has $n$ elements.  Then $M$ can be bounded as follows:
    \begin{align*} 
    M = \sum_{i=1}^{n}\ a_{i} &\leq n*a_{n} \\
    \mbox{and the number of required qubits} \\
    \NumBits{(n*a_{n})} &\leq \log_2{n} + \log_2{a_{n}}+1 \\
    & = O(\log (\max(n,a_{n})))
    \end{align*}
The number of required qubits for an SSP instance is bounded logarithmically, by the number of elements in $S$ and by its largest integer.  Such a result would normally be encouraging, but qubits are precious on quantum devices, so we investigated other approaches to reducing the necessary qubits.

\subsection{Varying-width set elements and sums}\label{sec:varwidth}
    While $M$ as computed above suffices to represent all values, an instance of SSP may specify integers in $S$ that require fewer qubits.  We therefore implemented a varying-width arithmetic package for SSP, so that each integer $a_i\in S$ occupies only the necessary \NumBits{a_{i}} qubits.  This in turn required redefining addition to accommodate inputs of differing qubit widths.

    Each shadow register $b_i$ is either a copy of its source register $a_i$, or it is~$0$, according to whether $x_i$ specifies that $a_i$ should be included in the sum.  Thus, each shadow register~$b_i$ is the same width as its source register~$a_i$.
    
    The additions performed in service of an instance of SSP on the shadow registers must be allocated sufficient qubits to hold their results.  With the shadow registers summed in some order (say, $b_1$ through $b_n$ for now), when $b_j$ is to be added to the growing partial sum, we allocate a register of width  \begin{equation} \NumBits{\sum_{i=1}^{j} b_{i}\label{eq:sum}}\end{equation}
    to hold the result.  The register must be this wide whether $x_i$ specifies $b_j$ to be the source value $a_j$ (so, included in the sum) or $0$ (so, excluded from the sum).  We therefore accommodate a sum at this point that potentially includes $a_j$.
    In other words, the sum at that point can be at most the sum of the integers in SSP that may have been incorporated thus far, and we allocate a result for the sum that accounts for the worst case.

Experiments in Section~\ref{sec:exper} show the effectiveness of properly sizing each integer from $S$ and of properly (worst-case) sizing the sums.
\subsection{Optimal ordering of integers}\label{sec:sorted}

$S$ is a set (or multiset, allowing duplicates) of integers, which ordinarily implies no order among its elements. However, Equation~\ref{eq:sum} shows that the number of bits at each summation step is related to the size of the integers and results that have been incorporated previously.
Modeled after the \emph{shortest job first} result~\cite{Cobham54}, we introduce and prove the following:
\begin{quote}
    \textit{Theorem} The partial sums developed in service of an SSP instance produce the smallest values when the integers of $S$ are considered in sorted order, ascending.
\end{quote}

\begin{proof}Consider a set of \( n \) integers, \( V = \{V_1, V_2, \ldots, V_n\} \). WLOG, assume that when sorted we obtain \[ V_1 \leq V_2 \leq \ldots \leq V_n \]
Our theorem is that every partial sum of the integers $\sum_{i=1}^{k} V_{i}$ is minimized if the integers are in sorted order.  By contradiction,
assume there exists an optimal schedule $\rho$ that minimizes the partial sum of each step in the addition and that $\rho$ is not sorted in ascending order. This means there must exist at least one pair of integers \( V_i \) and \( V_j \) where \( i < j \) in $\rho$ such that \( V_i > V_j \).

If \( V_i \) and \( V_j \) are swapped, the partial sum for all sums between \( V_j \) and \( V_i \) decreases by
\( V_i - V_j \), as \( V_j \) is added first. This swap does not affect the partial sums before \( V_j \) or after \( V_i \) in the sequence.

Thus, swapping \( V_i \) and \( V_j \) to follow ascending order strictly decreases the partial sums between the \( V_i \) and \( V_j \) additions. By applying this swap iteratively for all such inversions in $\rho$, $\rho$ becomes sorted in ascending order without increasing the partial sum at any step.

Since the assumption was $\rho$ is optimal but we can decrease the partial sums (or keep them the same) by enforcing ascending order, the initial assumption that $\rho$ is optimal must be incorrect. Therefore, sorting by ascending order minimizes the partial sum for all steps in the addition.\end{proof}

\medskip

\noindent Section~\ref{sec:exper} presents results on the savings realized by sorting set $S$.

\subsection{Uncomputing the carry bits}

Our library introduces new carry bits for each summation of qubits.  An addition of two qubit registers, each $m$~qubits wide, causes the allocation of $m$ carry bits to hold the values produced by each pair of bit additions.  The initial carry-in is $\ket{0}$ as shown in Figure~\ref{fig:onebitadd}, but we generate a carry-out at the leftmost addition that is never needed.

It is therefore plausible that $\frac{1}{3}$ of our qubits are allocated for carry operations.
It has been shown that a single carry bit could service all additions, threading its way through the circuit~\cite{cuccaro2004new}.  This would work as follows:
\begin{itemize}
    \item The carry-out computation is prepared as shown in Figure~\ref{fig:onebitadd} using three \CCNOT{} gates that target the carry-out bit.
    \item The value at that point is then used for the next two qubit additions as the carry-in. In Figure~\ref{fig:onebitadd}, the role of that qubit is the \emph{false} qubit shown at the top.  
    \item Once the control point is specified, the carry qubit can be \emph{uncomputed} by applying the inverse of the gates that gave it its current value.  The \CCNOT{} gate is its own inverse, so we simply replay the three operations shown in purple in Figure~\ref{fig:onebitadd} in reverse of the order in which they were applied.
    \item The carry qubit can now be used to receive carry-out of the next qubits' summation.
\end{itemize}
While this increases the gate count for the resulting circuit, we expect the number of qubits could be cut by $\frac{1}{3}$.  Another issue is the noise on the singleton carry qubit, relied upon throughout the entire circuit.  We consider this future work, to investigate the savings on qubits as compared with possible increased noise when so many operations are applied to a single qubit.

\subsection{Double-buffering of sums}\label{sec:dbuffer}

In the same way that the carry bit can be uncomputed and used for each carry computation~\cite{cuccaro2004new}, the partial sums that result from incorporating designated elements of $S$ could be accomplished using \emph{double buffering}.  The summation register that received its result at time $t$ could be used at time $t+2$, with the result from $t+1$ and the next integer from $S$ summed into the recycled register.  Instead of $n$ summation registers, we would need only~2.  The register from time $t$ must be \emph{uncomputed} so that it returns to its initial state and is ready to receive the summation at step $t+2$.

Using analysis similar to Section~\ref{sec:reducewidth}, we could save \[O(n\times \log(\max(n,a_{n})))\] qubits, where $n$ is the size of $S$ and $a_{n}$ is the largest integer in $S$.  This must be investigated because uncomputing has implications for noise and delay in the quantum circuit.
\section{Computing string equality}\label{sec:equality}

At the end of an SSP instance, the final sum developed by the various approaches described in Section~\ref{sec:qubits} must be compared with the target~$T$ for the SSP decision problem.  In this section, we present a new approach for computing equality between two strings (nominally considered to be the same width~$m$).  

Extensions to strings of different width are possible, but the implementation would depend on whether the strings are viewed as
\begin{itemize}
    \item numbers, in which case the shorter string can be prepended with sufficient $0$s to match the width of the longer string; or
    \item text, in which case the strings would be viewed as unequal owing to their difference in width.
\end{itemize}
\begin{figure}[htb]
\begin{center}
\includegraphics[scale=0.3]{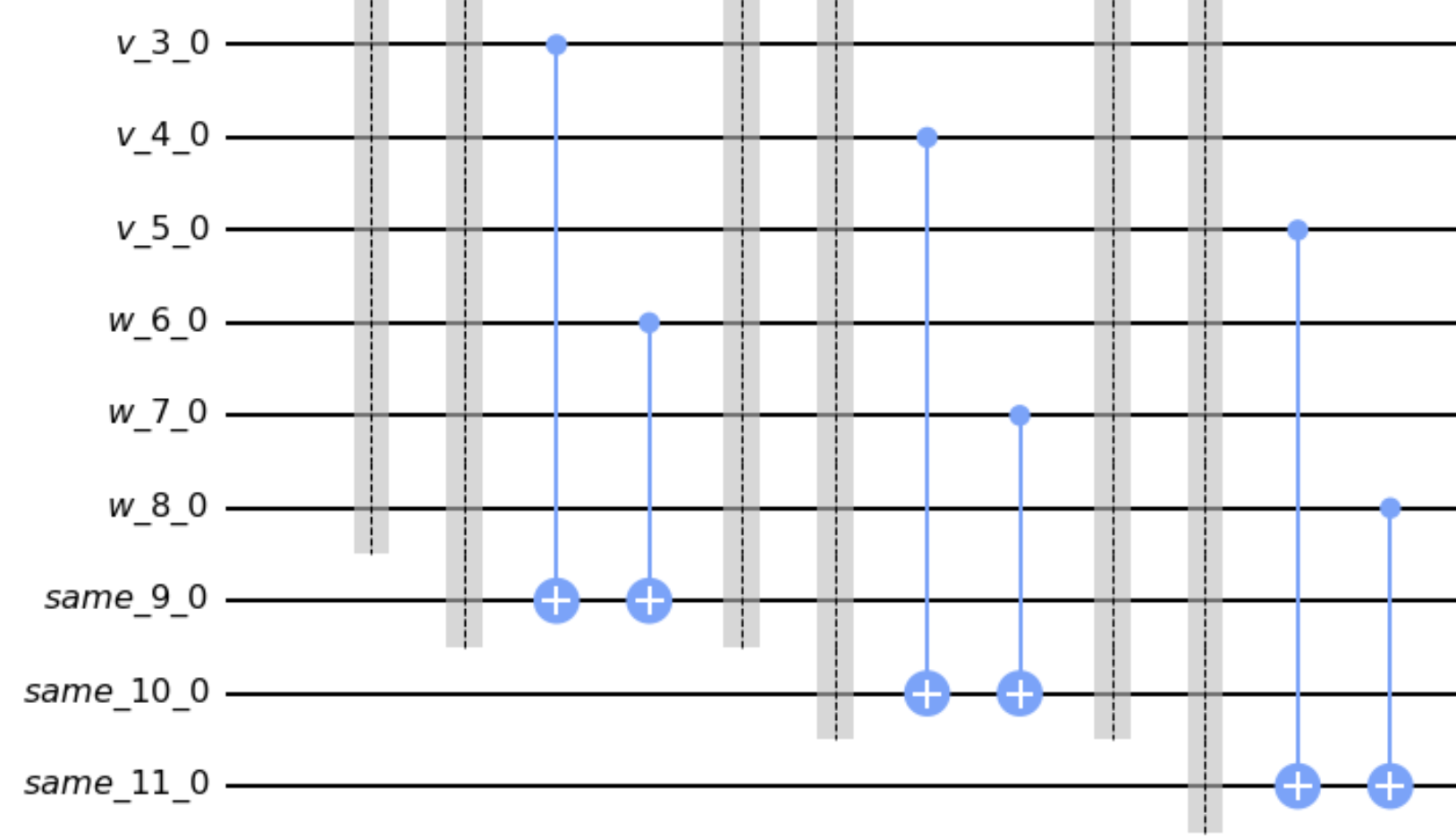} \\
(a) \\
\includegraphics[scale=0.4]{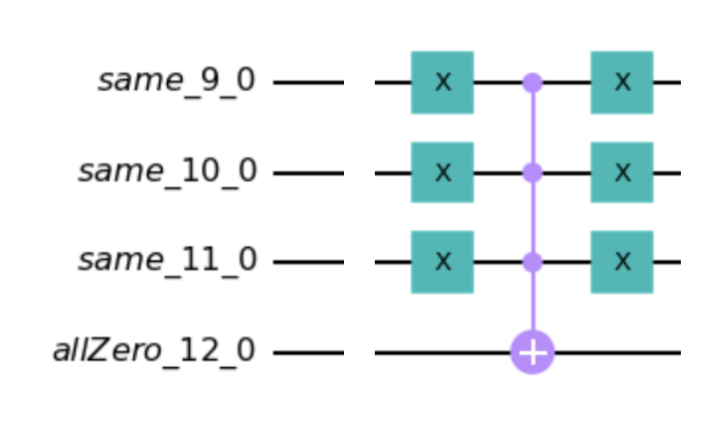} \\
(b)
\end{center}
\caption{Current method to determine $v=w$. In~(a), qubits $v_i$ and $w_i$ are compared via exclusive-or, with the resulting \texttt{same} bit $\ket{0}$ if and only if they agree.  The circuit continues in~(b), where the multi-controlled \X{} gate checks that all of the $same$ control bits are~0.}\label{fig:stdeql}
\end{figure}
The standard approach for a quantum circuit to detect whether strings $v$ and $w$ are identical is shown in Figure~\ref{fig:stdeql}.
\begin{enumerate}
    \item The equality of each pair of bits, $v^{i}$ and $w^{i}$ is computed by their exclusive-or.  Thus, we compute
    \[ same^{i} = v^{i} \oplus w^{i} \]
    using two \CNOT{} gates as shown in Figure~\ref{fig:stdeql}(a). The result $same^{i}$ is $0$ if and only if the $i^{\mbox{th}}$ bits of $v$ and $w$ are the same.
    The resulting string $same$ shows which qubit positions in $v$ and $w$ agree, and which do not.  Where agreement exists, the corresponding bit of $same$ is $\ket{0}$ 
    \item We next determine if \emph{all} the qubits in register $same$ are $\ket{0}$.  This is typically achieved using each as a control bit of a multi-controlled \X{} gate, whose target is some qubit $allZero$, as shown in Figure~\ref{fig:stdeql}(b).  Thus, $allZero=\ket{1}$ if and only if all the qubits of $same$ are $\ket{0}$. Because we are checking that all control bits are 0, each requires an \X{} gate prior to the multi-controlled \X{} gate (and after, possibly, to uncompute).
\end{enumerate}
We are concerned with the noise experienced for Grover on the IBM~Q devices~\cite{ibmq}, and the multi-control gates have been identified as a source of such noise~\cite{Wilson21}.
We therefore developed an alternate formulation for string equality.  We base our new approach on the following:
\[
\sum_{i=1}^{m} same^{i} = 0 \longleftrightarrow same=0
\]
In other words, $same$ is the integer~$0$ if and only if its digits sum to~$0$. The sum of such digits is at most~$m$, when each is a~$1$.  With reference to Equation~\ref{eq:sum}, we can represent the sum of~$m$ bits using \NumBits{m} bits.  For example, if $m=1000$ bits then its sum can be represented in $10$~bits.  That sum can in turn be represented in $4$ bits, and so on.  Each step reduces the number of bits needed logarithmically.  The number of such steps is thus $\Theta(\log\star(m))$.  In the limit, once we have a sum held in two bits, we can reduce the number of bits no further.  We finally test that the remaining two qubits are~$\ket{0}$ which takes a single \CCNOT{} gate.

We have thus reduced the problem of string equality to the problem of a series of additions, culminating in a~\CCNOT{} gate.  We perform $\Theta(\log\star(m))$ additions, each using only \CNOT{} and \CCNOT{} gates.   We have thus eliminated the need for a multi-controlled \CNOT{} gate with more than~2 control bits.

Investigation of this replacement for string equality is future work for us.  We intend to use IBM's emulators~\cite{ibmq} that can model the noise of their quantum computers.
\section{Oracle adaptation to allow SSP approximation}\label{sec:opt}

While the SSP decision problem is itself interesting, in terms of the hard problems we seek to solve, the \emph{optimization} form of the problem is more attractive:
\begin{quote}
    Given a (multi)set \[S = \{a_{1}, a_{2}, \ldots, a_{n}\}\] of $n$ positive integers, which subset of $S$ sums to a value as close to~$T$ as possible.
\end{quote}
We say ``which subset'' because any subset is a possible solution, including the empty set.  The optimization problem seeks a subset whose sum is as close to $T$ as possible given the integers in~$S$.

With an SSP instance formulated to find a subset whose sum is exactly $T$, a solution near $T$ could be found by repeated runs of a Grover search using targets acceptably close to~$T$.  However, the complexity of this approach introduces a linear factor related to the acceptable range around $T$.  Binary search cannot help if the oracle performs exact matches for its target. This is because the integer range around $T$ does not help to guide the search toward a solution.

In this section, we describe a simple adaptation of the oracle that allows for an approximate search around~$T$ in a single run of Grover's algorithm.  That run can also provide information about whether the approximate match is greater than or less than $T$.

This problem is certainly \NPH{} because the acceptably close range of $T$ could be just~$T$ itself.

\subsection{Approximate solution with a run}
The approach we have described in Section~\ref{sec:equality} culminates with an exact comparison with~$T$.  We describe here how that comparison can be relaxed to:
\begin{itemize}
    \item search for a sum whose value approximates $T$ and
    \item provide information for the next formulation of a Grover instance that will make the approximation more precise.
\end{itemize}
The approach works as follows, with an example shown in Figure~\ref{fig:approx}.
\begin{itemize}
    \item Within a given run of an SSP instance using Grover's algorithm, instead of matching against $T$ we match against all \emph{but} the least significant $k$ bits of $T$. 
    \item For $k=0$ this tests for an exact match against~$T$.  When viewed from right to left, as $k$ increases, the range for a match is correspondingly relaxed.
    \item In the quantum circuit, this is accomplished by matching only the most significant $n-k$ qubits and ignoring the others.  We can use the string-equality circuit described in Section~\ref{sec:equality}.
\end{itemize}
\def\BoxIt#1{\hbox to 1.2em{\hss #1\hss}}
\def\BoxItV#1{\BoxIt{\texttt{#1}}}
\begin{figure}[htb]
\begin{center}
\begin{tabular}{ccccccccccc}
\multicolumn{8}{c}{Value of $k$} \\
\BoxIt{8} & \BoxIt{7} & \BoxIt{6} & \BoxIt{5} & \BoxIt{4} & \BoxIt{3} & \BoxIt{2} & \BoxIt{1} & \BoxIt{0} \\\hline
\BoxItV{1} & \BoxItV{0} &
\BoxItV{0} & \BoxItV{1} &
\BoxItV{1} & \BoxItV{1} &
\BoxItV{0} & \BoxItV{1}
\end{tabular}
\end{center}
\caption{Example for approximate matching of $T$, with the bits of $T=157$ shown on the bottom.}\label{fig:approx}
\end{figure}
The effect of this approach, ignoring the least significant $k$ bits, is to search for match in the interval
\[
[T - (T \mod 2^{k}),\ T-(T \mod 2^{k}) + 2^{k}-1]
\]
The inclusive lower bound is obtained when the least significant $k$ bits are all~$0$, and the upper bound is obtained when they are all~$1$.
Using the example in Figure~\ref{fig:approx}, we obtain the following inclusive intervals for the approximate search, based on a particular provided value for $k$:
\begin{center}
    \begin{tabular}{ccccr}
    $k$ & $T \mod 2^{k}$ & Low & High &\multicolumn{1}{c}{Diff}\\\hline
    0 & 0 & 157 & 157 & 0\\
    1 & 1 & 156 & 157 & 1\\
    2 & 1 & 156 & 159 & 3\\
    3 & 5 & 152 & 159 & 7\\
    4 & 13 & 144 & 159 & 15\\
    5 & 29 & 128 & 159 & 31\\
    6 & 27 & 128 & 191 & 63\\
    7 & 27 & 128 & 255&  127
    \end{tabular}
\end{center}
The range expands at the low side or high side, depending on the bit pattern of $T$.  A zero at the current value of $k$ maintains the low value, and a one at that position maintains the high value. 
As shown in the rightmost column, the approximation becomes exponentially worse with each increase in $k$.

We use this idea to modify our oracle generator from Figure~\ref{fig:sketch} as follows:
\begin{itemize}
    \item The generator accepts a specific value for~$k$.
    \item The generated oracle determines success if $T$ is matched on all but the least significant $k$ bits.
\end{itemize}
Grover search then yields a superposition of subsets whose sums fall within the approximate range as described above for a given~$k$.  As described in Section~\ref{sec:grover}, a subset is represented as a particular value of~$x$ whose bits encode which elements of~$S$ are included in the subset whose sum falls within the approximate range.  The superposition is then measured to yield a particular value of $x$, which corresponds to a particular subset of $S$.

Once a subset is observed, subsequent Grover searches can be formulated to expand or contract the search space for a solution near~$T$.
\section{Experiments}\label{sec:exper}

To explore how effective our optimizations are on the number of qubits and gate operations of the circuit, we conducted empirical experiments on most of the ideas presented in Section~\ref{sec:qubits}.

These experiments were run on \texttt{Python}~3.11.4 and \texttt{Qiskit}~0.45. A record of the raw data and mean calculations is archived in our \texttt{GitHub} repository.\footnote{\texttt{https://github.com/ABenoit0226/quantum-place-route}} 

\subsection{Setup}
For testing, we generated instances of the Subset Sum problem with a randomized (multi)set of positive integers and a target between 1 and the sum of the entire set: an instance with a target beyond the entire set's sum clearly has no solution.  We vary the maximum value allowed in the set between 64, 128, and 256. Our primary goal was to reduce the qubit requirements of the oracle, but we measured both qubit and gate operations as a function of the number of values in the set. 

The number of qubits used in each circuit was calculated by counting each register created using our Quantum Circuit (QC) class. The gates of interest were counted by subtracting the uses of cosmetic quantum circuit barriers and measurement operations from a built-in \texttt{Qiskit} method for counting the number of circuit operations.

To obtain meaningful measurements, we compute the mean number of qubits and gate operations used over 100 runs (each with a randomized SSP instance) of each problem. Experiments were run between set sizes of~5 and~100 in increments of~5.

We evaluate 4 configurations of the subset sum implementation to demonstrate our circuit optimizations:
\begin{enumerate}
  \item The baseline configuration uses fixed-width, 32-bit registers to hold integers and perform addition operations.
  \item Here, we apply the optimization described in Section~\ref{sec:varwidth} and use varying-width registers, but only for the set's elements.  We continue to use fixed-width registers for the partial sums based on the size of the total sum of all elements in the set, as described in Section~\ref{sec:reducewidth}.
  \item Continuing with the optimization described in Section~\ref{sec:varwidth}, we used varying-width registers for set elements \emph{and} the partial sums.
  \item Finally, we use the above optimizations but also perform sums using the sorted order of the set~$S$ as described in Section~\ref{sec:sorted}.
\end{enumerate}

\subsection{Qubit Results}

\begin{figure}[htb]
    \begin{center}
    \includegraphics[width=.4\textwidth]{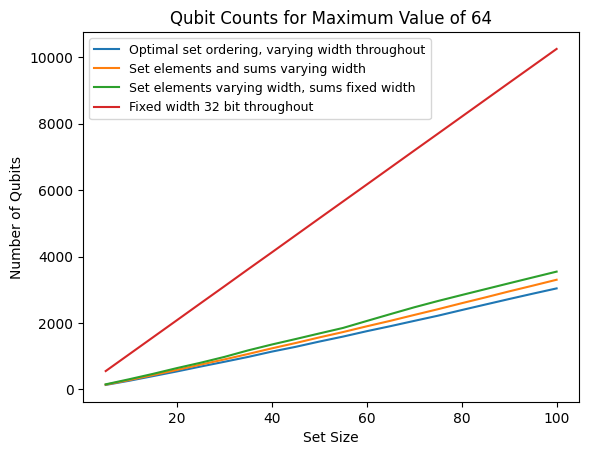}
    \includegraphics[width=.4\textwidth]{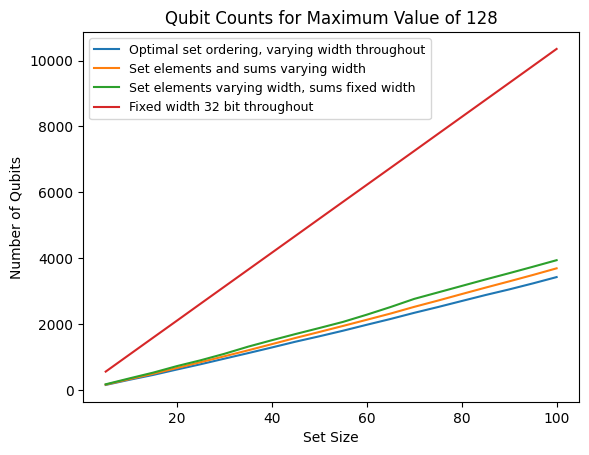}
    \includegraphics[width=.4\textwidth]{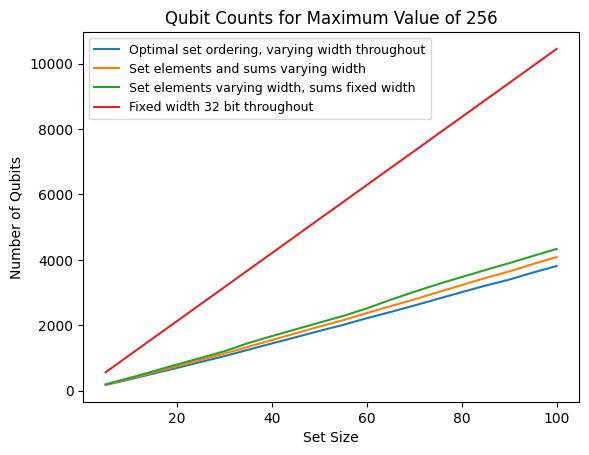}
    \end{center}
    \caption{Mean qubit usage of optimization vs. set size}
    \label{fig:meanqubits}

\end{figure}

As shown in Figure~\ref{fig:meanqubits}, our transition to using varying width integers rather than a standard 32-bit integer (as a classical computer does) reduces the number of qubits used to create the SSP solver circuit substantially. This is particularly noticeable as the size of the problem grows and the qubit count between the baseline and the optimization diverges. This value manifests as a reduction of  $\sim$$58-73\%$ (see Figure~\ref{fig:tablequbits}).

We can further reduce the number of qubits used through two other optimizations in our calculation of the partial sums: using varying width partial sums and ordering the set before summation (see Section II-D). These optimizations further reduce the number of qubits used by $\sim$$1.5-2.5\%$ each (see Figure~\ref{fig:tablequbits}).

As the maximum value allowed in the set increases, the percentage reduction in qubits decreases. This is expected as the baseline uses 32-bit registers by default and larger integers use a register size closer to~32 so the varying width methods save fewer qubits.

The decrease in qubit reduction as the set size increases is comes from the fact that the optimized methods' functions are respectively less linear. Despite appearing linear in terms of qubits/set size (see Figure~\ref{fig:meanqubits}), the fully optimized method's function actually increases in slope by roughly 15\% between 10 and 100 values, whereas the fixed method's function experiences a negligible change ($<$.01\%) when the maximum allowed value is 256. This is due to partial sums approaching 32 bit width as the set size grows.

\begin{figure}[htb]

    \begin{center}
    \textbf{Max Value = 64}\\[0.5em]
    \begin{tabular}{|c|c|c|c|}
    \hline
    Set size: & 5 values & 50 values & 100 values \\
    \hline
         Sums fixed width & $72.80 \%$ & $67.27 \%$  & $65.40 \%$ \\
    \hline
         Sums varying width & $74.57 \%$  & $69.59 \%$ & $67.79 \%$ \\
    \hline
         Optimal ordering & $75.96 \%$  & $72.02 \%$ & $70.33 \%$ \\
    \hline
    \end{tabular}
    \end{center}

    \begin{center}
    \textbf{Max Value = 128} \\[0.5em]
    \begin{tabular}{|c|c|c|c|}
    \hline
    Set size: & 5 values & 50 values & 100 values \\
    \hline
   
         Sums fixed width & $69.13 \%$ & $63.83 \%$  & $61.95 \%$ \\
    \hline
         Sums varying width & $71.02 \%$  & $66.12 \%$ & $64.33 \%$ \\
    \hline
         Optimal ordering & $72.39 \%$  & $68.66 \%$ & $66.90 \%$ \\
    \hline
    
    \end{tabular}
    
    \end{center}

    \begin{center}
    \textbf{Max Value = 256}\\[0.5em]
    \begin{tabular}{|c|c|c|c|}
    \hline
    Set size: & 5 values & 50 values & 100 values \\
    \hline
         Sums fixed width & $65.67 \%$ & $60.37 \%$  & $58.53 \%$ \\
    \hline
         Sums varying width & $67.43 \%$  & $62.68 \%$ & $60.91 \%$ \\
    \hline
         Optimal ordering & $68.92 \%$  & $65.19 \%$ & $63.53 \%$ \\
    \hline

    \end{tabular}
    \end{center}
    \caption{Percentage of qubits saved over Fixed 32 bit by optimization (max value in set = 64, 128, 256 respectively)}
    \label{fig:tablequbits}
\end{figure}

\subsection{Gate Results}
\begin{figure}[htb]
    \begin{center}
    \includegraphics[width=.4\textwidth]{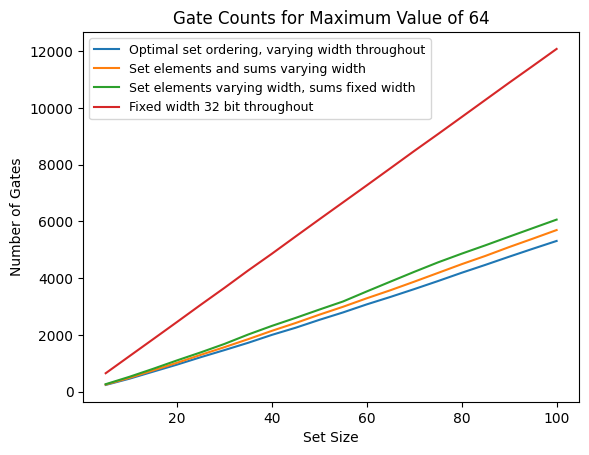}
    \label{fig:enter-label}
    \includegraphics[width=.4\textwidth]{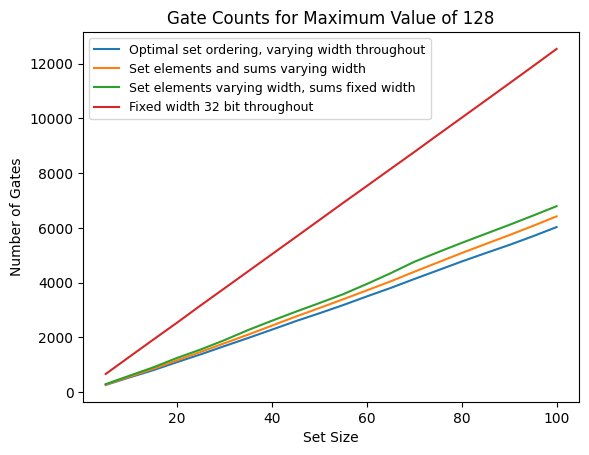}
    \includegraphics[width=.4\textwidth]{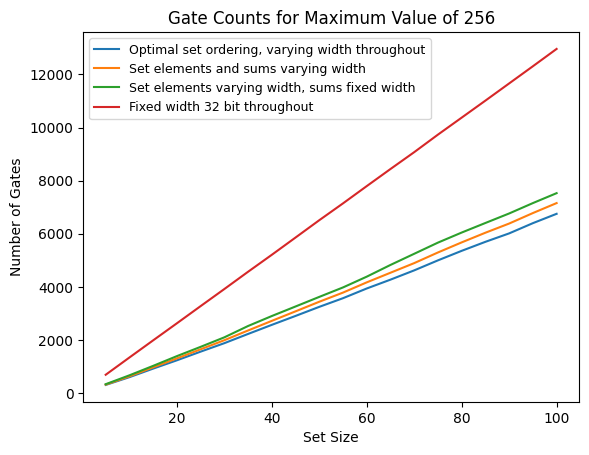}
    \caption{Mean gate usage of optimization vs. set size}\label{fig:meangates}
    \end{center}
\end{figure}

    Reducing the number of gate operations is also important, as each gate introduces noise and delay in the circuit. As shown in Figure~\ref{fig:meangates}, using a width based on the total sum of all set elements instead of 32 bits significantly reduces the number of gates used in the SSP circuit as well. This value is further optimized by using varying widths for the partial sums and by ordering the set. As the problem grows, the gate count between the baseline and optimization diverges, although to a lesser extent than with the qubit counts. Overall, we see a reduction of $\sim$$42-62\%$ from fixed-width to the size of the total sum of all elements, an additional $\sim$$2.5-3\%$ by using varying width in the partial sums, and another $\sim$$1.5-3.25\%$ from ordering the elements before summing.

    Similar to the qubit results, there are slightly diminished reductions from the optimizations compared to the baseline as the maximum value allowed in the set increases and approaches 32 bits in size. There are also diminished reductions for more values in the set as the resulting sums are closer to the size of the 32-bit sums.

\begin{figure}[htb]

    \begin{center}
    \textbf{Max Value = 64} \\[0.5em]
    \begin{tabular}{|c|c|c|c|}
    \hline
    
    Set size: & 5 values & 50 values & 100 values \\
    \hline
         Sums fixed width & $61.73 \%$ & $52.42 \%$  & $49.82 \%$ \\
    \hline
         Sums varying width & $64.19 \%$  & $55.38 \%$ & $52.87 \%$ \\
    \hline
         Optimal ordering& $65.81 \%$  & $58.38 \%$ & $56.05 \%$ \\
    \hline
    \end{tabular}
    \end{center}

    \begin{center}
    \textbf{Max Value = 128}\\[0.5em]
    \begin{tabular}{|c|c|c|c|}
    \hline
    
    Set size: & 5 values & 50 values & 100 values \\
    \hline
         Sums fixed width & $57.35 \%$ & $48.27 \%$  & $45.82 \%$ \\
    \hline
         Sums varying width & $59.91 \%$  & $51.13 \%$ & $48.78 \%$ \\
    \hline
         Optimal ordering& $61.47 \%$  & $54.19 \%$ & $51.91 \%$ \\
    \hline
    
    \end{tabular}
    
    \end{center}

    \begin{center}
    \textbf{Max Value = 256}\\[0.5em]
    \begin{tabular}{|c|c|c|c|}
    \hline
    Set size: & 5 values & 50 values & 100 values \\
    \hline
         Sums fixed width & $53.53\%$ & $44.38 \%$  & $41.92 \%$ \\
    \hline
         Sums varying width & $55.85 \%$  & $47.19 \%$ & $44.81 \%$ \\
    \hline
         Optimal ordering & $57.52 \%$  & $50.14 \%$ & $47.92 \%$ \\
    \hline

    \end{tabular}
    \end{center}
    \caption{Percentage of gates saved over Fixed 32 bit by optimization (max value in set = 64, 128, 256 respectively)}
    \label{fig:tablegates}
\end{figure}


\section{Oracle library}\label{sec:library}

Our library for initializing and executing subset sum oracles consists of 3 classes that support quantum circuits, arithmetic, and the Subset Sum problem.  We also developed methods for testing and data generation that produced the data reported in Section~\ref{sec:exper}.

\subsection{Quantum Circuit Class}
We define a base quantum circuit class, \texttt{QC}, a wrapper class for the \texttt{Qiskit} Quantum Circuit class. It creates and manages the quantum circuit, with methods for adding registers that allow us to account for qubit usage and name the registers according to their use as an aid to understanding the generated circuit.  The class provides for all basic Boolean operations and offers methods that optimize the generation of carry bits and chains of exclusive-ors, with a resulting circuit shown in Figure~\ref{fig:onebitadd}.  While those operations can in theory be generated from the basic Boolean operators, qubits are spared by realizing these operations in a more compact form.  For example, a direct realization of exclusive-or in terms of the more basic Boolean operations burns 6~qubits for temporary values and uses  10~quantum gates. 
 We do not report experimental results separately for these improvements but use them throughout our experiments.  

 The methods in this and the following classes accept information to perform a given operation and can either
 \begin{itemize}
     \item generate a new register to contain the result, or
     \item accept a register as a parameter that should receive the result.
 \end{itemize}
 The first use case is \emph{functional} in nature and allowed us to write complicated operations very easily in terms of their constituent operations.


\subsection{Quantum Arithmetic Class}
Following the \emph{decorator} pattern~\cite{gangoffour}, we developed a
class \texttt{QVarArith} that decorates the \texttt{QC} class.  This supports the instantiation of integers, arithmetic on integer values, and comparison of integer values.

The class supports varying-width arithmetic by default, but it can be invoked to create or add integers of a specified width.  This allows us to experiment with the savings of varying-width arithmetic as described in Section~\ref{sec:varwidth}.

\subsection{Subset Sum Class}
We define a class to manage instances of the Subset Sum problem, \texttt{QSubsetSum}. The class takes a list of integer values (our set $S$), and an integer target ($T$) generates a \texttt{QC} object containing the oracle.

\section{Conclusion and future work}\label{sec:conclu}
We have examined several techniques for reducing the number of qubits needed to generate an oracle for the Subset Sum problem. We presented results that show the greatest savings in qubits were seen by moving from fixed-width arithmetic to varying-width for the integers in~$S$.  Extending that idea to the sums brought some improvement, and sorting the set to minimize the partial sums yielded additional, modest gains.  Our circuits also required fewer gates with each optimization, but qubits had been the limiting factor for running the subset sum problem in emulation or on actual hardware.

We described a new technique for bit-string comparisons in a quantum circuit that avoids a multi-controlled gate.  We described a method for adapting the oracle to allow an approximate search.

The code producing our results in Section~\ref{sec:exper} is available in the \texttt{GitHub} repository referenced there. For future work we are interested in the following directions:
\begin{itemize}
    \item We hope to apply the idea from~\cite{cuccaro2004new} to save carry bits by uncomputing and reusing them.  An issue could be the accumulation of noise using just a single qubit for such a purpose.  We hope to consider a qubit uncompute and noise management scheme, in which we could use a given qubit for carries until the noise might become intolerable, prior to which point we would allocate a new qubit for carries.
    \item The double-buffering idea described in Section~\ref{sec:dbuffer} could significantly reduce the qubits needed for the oracle, at the expense of a deeper circuit.  The depth matters in terms of delay and noise.  We could manage those buffers in the same manner as the carry qubit, described above.  We would reuse a given pair of sum registers until noise or depth becomes egregious.
    \item Our emulations using Grover's search provide great results, but the same circuits on current IBM hardware~\cite{ibmq} are disappointing.  We hope to localize the source of noise, which we believe to be the multi-controlled \X{} gates.  The technique in Section~\ref{sec:equality} avoids one such gate, but the diffuser step of Grover's algorithm still uses such a gate.
\end{itemize}


\section*{Acknowledgement}

The authors thank Roger Chamberlain for conversations about hard optimization problems.



%


\bibliographystyle{IEEEtran}
\bibliography{IEEEabrv,refs}

\end{document}